\documentclass[a4paper]{article}

\usepackage{INTERSPEECH2020}

\usepackage{graphicx}
\usepackage{float} 
\usepackage{subfig}

\usepackage{multirow}
\usepackage{caption}
\usepackage{cite}
\usepackage{array}
\usepackage{url}

\title{Incorporating Broad Phonetic Information for Speech Enhancement}
\name{Yen-Ju Lu$^1$, Chien-Feng Liao$^1$, Xugang Lu$^2$, Jeih-weih Hung$^3$, Yu Tsao$^1$}
\address{
  $^1$Research Center for Information Technology Innovation, Academic Sinica, Taiwan \\
  $^2$National Institute of Information and Communications Technology, Japan\\
  $^3$National Chi Nan University, Taiwan}
\email{\{r03942063, r06946002\}@ntu.edu.tw, xugang.lu@nict.go.jp, jwhung@ncnu.edu.tw, yu.tsao@citi.sinica.edu.tw }

\begin{document}

\maketitle
\begin{abstract}
In noisy conditions, knowing speech contents facilitates listeners to more effectively suppress background noise components and to retrieve pure speech signals. Previous studies have also confirmed the benefits of incorporating phonetic information in a speech enhancement (SE) system to achieve better denoising performance. To obtain the phonetic information, we usually prepare a phoneme-based acoustic model, which is trained using speech waveforms and phoneme labels. Despite performing well in normal noisy conditions, when operating in very noisy conditions, however, the recognized phonemes may be erroneous and thus misguide the SE process. To overcome the limitation, this study proposes to incorporate the broad phonetic class (BPC) information into the SE process. We have investigated three criteria to build the BPC, including two knowledge-based criteria: place and manner of articulatory and one data-driven criterion. Moreover, the recognition accuracies of BPCs are much higher than that of phonemes, thus providing more accurate phonetic information to guide the SE process under very noisy conditions. Experimental results demonstrate that the proposed SE with the BPC information framework can achieve notable performance improvements over the baseline system and an SE system using monophonic information in terms of both speech quality intelligibility on the TIMIT dataset. 
\end{abstract}

\section{Introduction}
Speech enhancement (SE) systems aim to transform the distorted speech signal to an enhanced one with improved speech intelligibility and quality. In many real-world applications, such as speech coding \cite{martin1999new,accardi1999modular}, assistive listening devices \cite{wang2017deep,lai2016deep} and automatic speech recognition (ASR) \cite{li2014overview, erdogan2015phase, weninger2015speech}, SE has been widely used as a front-end processor, which effectively reduces noise components and improves the overall performance. 
Recently, along with the burgeoning of deep learning (DL), applying DL-based models for the SE task has been extensively investigated \cite{wang2014training,xia2014wiener,lu2013speech, xu2014regression,kolbaek2016speech,tan2019real,Qi2019SE}. In particular, the DL-based models provide a strong capability of modeling non-linear and complex transformations from noisy and clean speech, and therefore the DL-based SE approaches have yielded notable improvements over traditional SE methods, especially under very low signal-to-noise-ratio (SNR) and non-stationary noise conditions. Another vital feature of DL-based models is their high flexibility to incorporate heterogeneous data, which may not be easily performed for traditional signal processing-based approaches. Previous studies have confirmed that the face/lip images \cite{8323326} and symbolic sequences for acoustic signals \cite{Liao2019} can be incorporated into an SE system.

The phonetic information generated from an acoustic model (AM) has also been adopted to guide the SE process and achieve improved denoising performance. In \cite{chen2015speech, 7403942, 7178797}, an AM and an SE system are trained jointly, where one system’s input depends on the other’s output. In \cite{9054334}, the SE result was treated as an adaptive feature to generate a phoneme posteriorgram (PPG) more precisely. Although achieving improved performance under ordinary conditions, the AM may generate inaccurate PPGs under very noisy or acoustically mismatched conditions, which may misguide the SE process to generate even poorer results.
To overcome the issue, we proposed to incorporate the broad phonetic class (BPC) posteriorgram (BPPG) in the SE system. We argue that the speech signals within the same BPC share the same noisy-to-clean transformation, and the BPPG can guide the SE process well even under very noisy conditions. The main concept of BPC is clustering phonemes with similar characteristics into a broad class. The criteria for clustering can be either knowledge-based or data-driven. The most common knowledge-based criteria exploit the manner and place of articulations for pronouncing the phonemes \cite{ladefoged2014course}. For data-driving criteria, the confusions between phonemes are measured based on some predefined evaluation metrics, and the phonemes that generate the closest metric scores (easily confused) are clustered into the same class. As compared to phoneme-based AM, the labels needed to train the BPC-AM are much more easily accessed, and the recognition accuracy is generally higher, especially in noisy conditions. In the past, the BPC has been widely used in various speech-related applications, including speaking rate estimation \cite{yuan2010robust}, multilingual systems \cite{vzgank2005data} and phone recognition \cite{tsao2005study, morris2008conditional, lee2019multi, scanlon2007using}. Experimental results have confirmed that BPCs can effectively boost recognition rates as a pre-processor or in unsupervised learning tasks. 

In this paper, we proposed to combine the BPPG into an SE system, termed BPSE. Three criteria that we used to define the BPC are tested. They are based on manner of articulation, place of articulations, and data-driven, thus termed BPC(M), BPC(P) and BPC(D). We evaluated the proposed BPSE on the TIMIT dataset. Two standard evaluation metrics, speech quality (PESQ) \cite{rix2001perceptual} and short-time objective intelligibility (STOI) \cite{taal2011algorithm}, are used to  evaluate the SE performance. For comparison, we implemented an SE system with monophonic PPG and two baseline SE systems. Experimental results first validate the effectiveness of incorporating the BPC information for the SE system. These results also confirm that the BPSE system outperforms an SE with monophonic PPG, especially under very noisy conditions, where PPG may be inaccurate.

The rest of this paper is organized as follows. Section 2 introduces the criteria used to define BPC. Section 3 details the proposed BPSE system. Section 4 presents the experimental setup and results. Finally, Section 5 concludes this paper.

\section{Broad Phonetic Classes}
In this section, we introduce two knowledge-based and one data-driven criteria used for defining BPC.

\subsection{Knowledge-based criterion}
In the phonetic field, consonants can be readily classified based on the manner and place where the vocal tract obstructs the airstream. Accordingly, the entire consonants can be divided into different place/manner-wise articulation classes. Distinct characteristics are found between the different manners/places of articulation, and different manner can even be easily revealed by the shape of waveform \cite{ladefoged2014course}.
It is also found in \cite{scanlon2007using} that phonemes belonging to the same manner/place of articulation contain very similar spectral characteristics, and may generate high confusion results when performing speech recognition. 

In this study, based on the manner of articulation, we cluster all the 60 phonemes into 5 clusters: vowels, stops, fricatives, nasals and silence as suggested in \cite{scanlon2007using}, where the diphthongs and semi-vowels are merged into the vowel class. On the other hand, according to the place of articulation, we divide phonemes into 9 clusters: bilabial, labiodental, dental, alveolar, postalveolar, velar, glottal, vowels, and silence, as suggested in \cite{ladefoged2014course}. Please note that all the vowels are clustered into one distinct class in both manner and place articulations. The underlying reason is that these two classification criteria are connected with the manner/place where the vocal tract obstructs the airstream, while the vowels do not have such properties.

\subsection{Data-driven criterion}
In contrast to the knowledge-based criteria, the data-driven criterion conducts phoneme clustering through the phoneme similarity measured by an ASR. Based on \cite{lopes2012broad}, the confusion matrix, $\textbf{M}$, contains information about the similarities between each pair of phonemes, where the entry $\textbf{M}_{ij}$ denotes the number of the event for phoneme $i$ being mistakenly recognized as phoneme $j$. A symmetric similarity matrix, $\textbf{S}$, can be computed from the confusion matrix, $\textbf{M}$, where $\textbf{S}_{ij}$, the similarity between phonemes $i$ and $j$, is computed by  (\ref{Houtgast}):

\begin{equation}
  \textbf{S}_{ij} = \textbf{S}_{ji} = \sum_{k\neq i, j} \mathrm{min}(\textbf{M}_{ik},\textbf{M}_{jk})
  \label{Houtgast}
\end{equation}
where $k$ is the phone index and $k \neq i, j$. From Eq. (\ref{Houtgast}), we note that each entry in the similarity matrix presents the distance between a pair of phonemes based on how often they are confused by the ASR results. When applying the similarity metric to cluster phonemes, we initially define each phoneme as one distinct cluster. Then, we gradually reduce the cluster number by merging the nearest clusters. This process repeats until the cluster number meets our expectations. In Table ~\ref{tab:confusion}, we list the clustering results (9 clusters, which are recommended in \cite{lopes2012broad}) obtained by the data-driven criterion on the TIMIT dataset.

\begin{table}[t!]
\caption{Clustering results by data-driven on TIMIT.}
\label{tab:confusion}
\centering
\begin{tabular}{ p{2cm}  p{5cm} } 
 \hline
 Clusters & TIMIT phoneme label \\
 \hline
Cluster1& bcl, dcl, epi, gcl, kcl, pau, pcl, q, tcl \\
Cluster2& b, d, dh, f, g, k, p, t, th, v \\
Cluster3& y \\
Cluster4& hh, hv \\
Cluster5& dx, em, en, m, n, ng, nx \\
Cluster6& aa, ae, ah, ao, aw, ax, ax-h, axr, ay, eh, el, er, ey, ih, ix, iy, l, ow, oy, r, uh, uw, ux, w \\
Cluster7& ch, jh, s, sh, z, zh \\
Cluster8& eng \\
Cluster9 & h\# \\
 \hline
\end{tabular}
\vspace{-4mm}
\end{table}

\begin{figure}
 \centering
 \includegraphics[width=\linewidth]{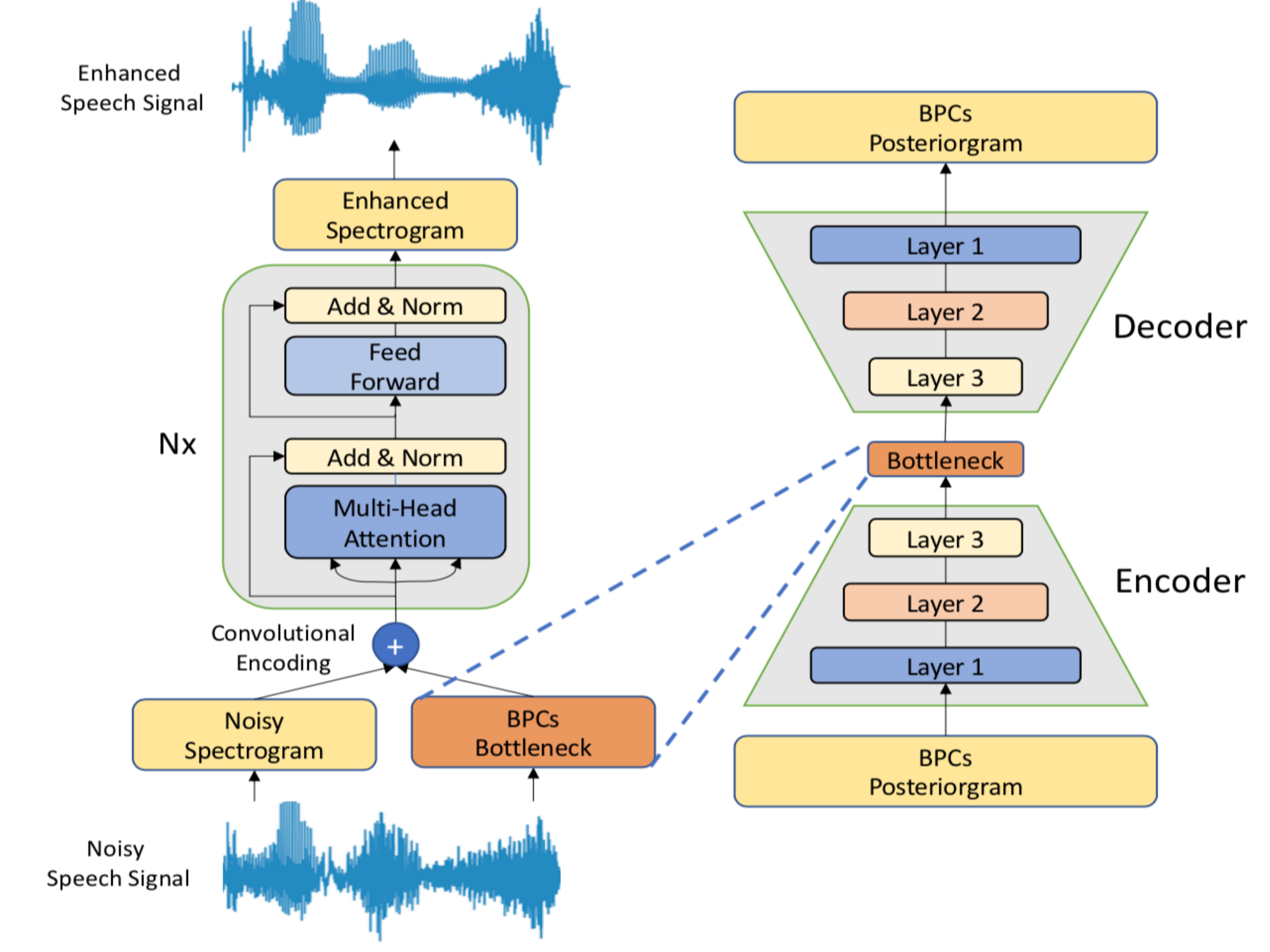}
 \caption{The architecture of the proposed BPSE system} 
 \label{fig:SE system}
\end{figure}


\section{Proposed BPSE system}
Figure \ref{fig:SE system} shows the overall architecture of the proposed BPSE system, which consists of a Transformer-based SE model \cite{vaswani2017attention} and a BPPG extractor.

\subsection{PPG and BPPG extractors}
To extract PPG and BPPG, we trained acoustic models (AMs) with noisy speech data as input, and phoneme labels and BPC labels as output, respectively. All AMs were trained via an off-the-shelf Kaldi recipe \cite{povey2011kaldi} with the same DNN-HMM architecture; the DNN has 7 hidden layers, each layer containing 1024 neurons. With the trained AM and given a speech utterance as the respective input, we collected the sequence of triphone state posteriors to form the PPG and BPPG. As for the case of phone unit, since the total number of triphone states is huge, we used the mono-phone state PPG instead. In order to keep the vector dimensions of PPG and BPPG to be moderate, an auto-encoder (AE) model, as shown in the right part of Figure \ref{fig:SE system}, was used to reduce the size of the BPPG vector. In particular, the input and output of the AE model are both BPPG vectors, and the latent representations with a dimension of 96 were used as the final BPPG features. Finally, the BPPG features are appended to the noisy spectrogram vectors and then served as the input to the SE system. 

\subsection{Transformer-based SE Network}
Transformer is an attention-based deep neural network, originally proposed for machine translation \cite{vaswani2017attention} and later explored in many other natural language processing tasks, exhibiting stark improvements over the well-known recurrent neural networks (RNNs). Recently, it was further exploited in the SE task \cite{9053591}, in which the Transformer model was also shown to outperform convolutional-based and recurrent-based models. The self-attention mechanism of Transformer has been well studied in the SE literature \cite{hao2019attention, 9053288, 9053214}, and one of its particularities is to allow the model to learn long-range dependencies within the sequence efficiently. Also, since Transformer can process the whole sequence in parallel, the respective computational time is reduced relative to RNNs.

The original Transformer consists of encoder and decoder networks for sequence-to-sequence learning. In our method, the decoder part is omitted since the input and output sequences have the same length during the SE process. Another modification is that we use convolutional layers to replace positional encoding in order to inject relative location information to the frames in the sequence, which is distinct from \cite{9053591}. The causal implementation on the transformer was applied for the real-time processing scenario. The rest of the architecture is implemented as a standard Transformer shown in the left part of Figure \ref{fig:SE system}, which is composed of $N$ attention blocks. In each attention block, the first sub-layer is the multi-head self-attention (MHSA), followed by a feed-forward network containing two fully-connected layers. Both sub-layers are followed by a residual connection to the input and a layer normalization \cite{ba2016layer}. Finally, the Transformer output is projected back to the frequency dimension using a fully-connected layer with ReLU activation, and the corresponding mean-absolute-error relative to the clean speech is computed.


\begin{figure}[tp!]

\subfloat[Spectrogram and recognition result at 10dB SNR ]{%
  \includegraphics[clip,width=\columnwidth]{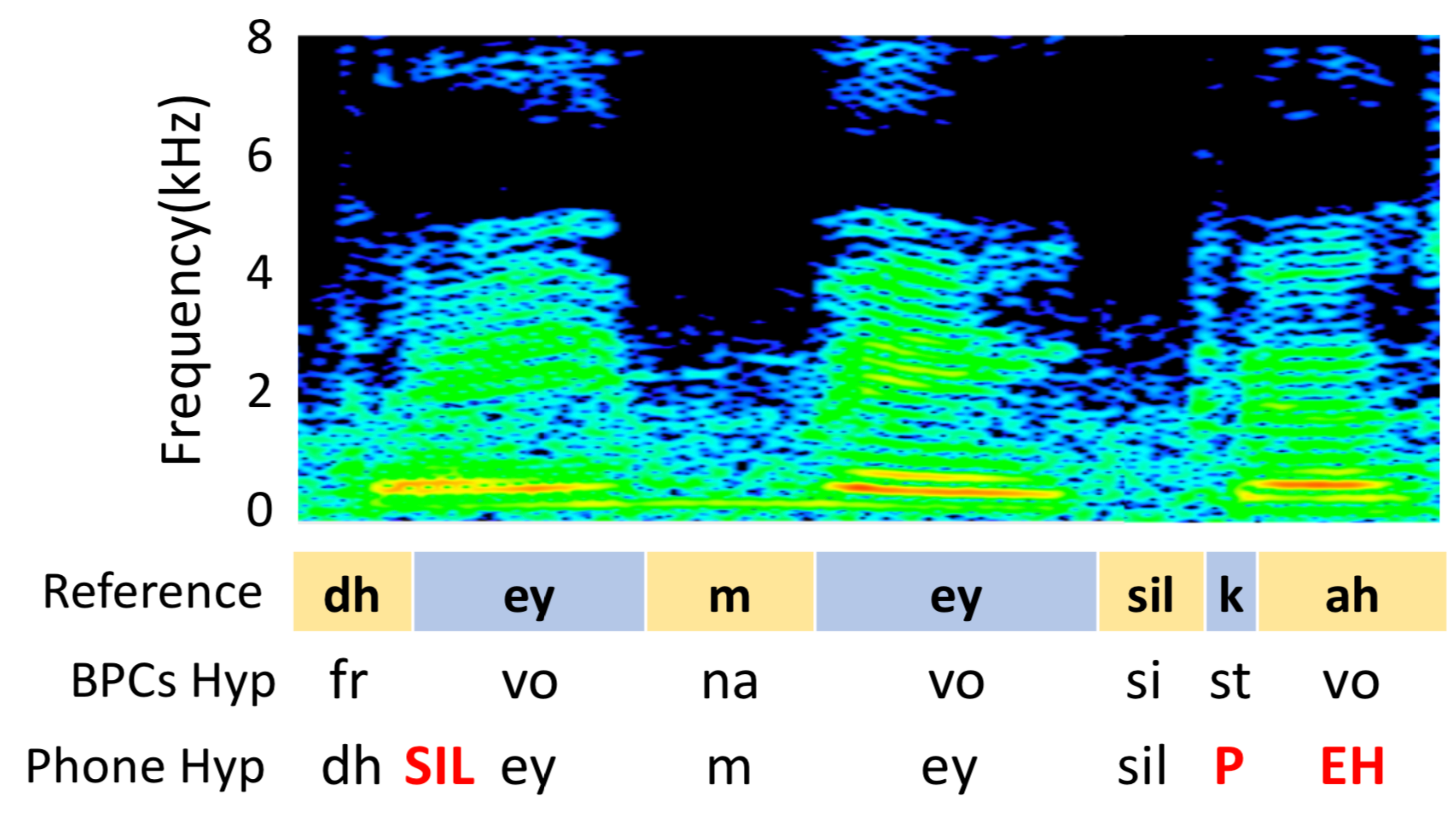}%
  \label{fig:recognition_10dB}
}

\subfloat[Spectrogram and recognition result at 0dB SNR level]{%
  \includegraphics[clip,width=\columnwidth]{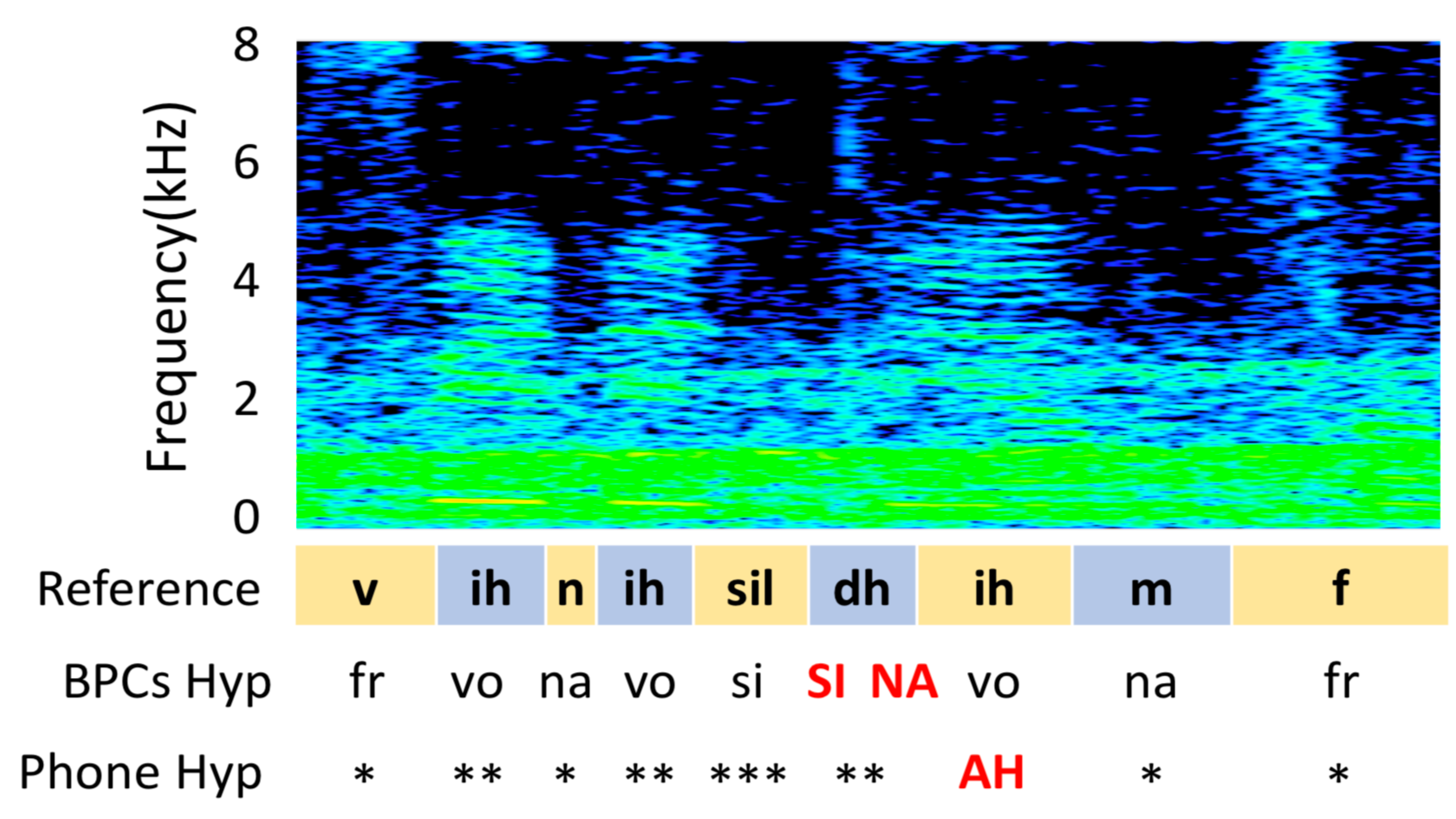}%
  \label{fig:recognition_0dB}
}

\caption{Monophonic and BPC recognition results of a sample utterance in two SNR levels. The length of slots in Reference denotes the start/end time. Phone Hyp and BPC Hyp mean the recognition results by phone-based and BPC-based AMs, respectively. The BPC labels include vowels(vo), stops(st), fricatives(fr), nasals(na) and silence(si). The uppercase phone/BPC labels represent recognition errors, including both insertions and substitutions. The star labels represent deletions.}
\label{fig:recognition}
\vspace{-4mm}
\end{figure}


\section{Experiment}
\subsection{Experimental setup}
The experiments were conducted using the TIMIT database \cite{garofolo1993timit} together with other noise sources \cite{hu2004100} for both AM and SE model. The 3696 utterances from the TIMIT training set (excluding SA files) were used and randomly corrupted with 100 noise types from \cite{hu2004100} at 32 different SNR levels, amounting to 3696 noisy-clean paired training utterances. The test set was constructed by adding 5 unseen noise types into all of the core test set of the TIMIT database (including 192 utterances) at 5 SNR levels (15 dB, 10 dB, 5 dB, 0 dB  and -5 dB). The speakers in the test set are different from those in the training set.

The speech waveforms were recorded at 16 kHz sampling rate. The short-time Fourier transform (STFT) with a Hamming window size of 32 ms and a hop size of 16 ms was applied to convert the speech waveforms into spectral features, each with 257 dimensions. The $log1p$ function ($log1p(x)=log(1+x)$) was adopted on the magnitude spectrogram to enforce the range of each coefficient to be positive. During training, a segment of 64 frames was used as the input unit to the SE system. During testing, the enhanced spectral features were synthesized back to the waveform signals via the inverse STFT and an overlap-add procedure. The phases of the noisy signals were used for the waveform reconstruction. The Adam optimizer \cite{kingma2014adam} was used to set the learning rate, and an early stopping was performed according to the validation set to prevent overfitting.

The autoencoder consisted of three fully connected layers in the encoder network, with the dimensions of $[512, 256, 96]$, and the decoder was symmetric to the encoder. The 96-dimensional bottleneck feature from the encoder was processed by a sigmoid function to limit the value range and further concatenated with the noisy spectrogram. The concatenated features were then fed into the SE module. For the Transformer, four 1-D convolutional layers were used to encode the input feature. The channel sizes were $[1024, 512, 256, 128]$. The filter size and stride were set to 3 and 1, respectively. We set the number of attention blocks $N$ to 8. The MHSA consisted of 8 heads and 64 neurons for each head. The two layers of the feed-forward network consisted of 512 and 256 neurons, respectively. Leaky ReLU was used as the activation function. 

Based on the model architecture shown in Fig. \ref{fig:SE system}, we implemented three systems using three types of BPPG, namely BPPG(M), BPPG(P), and BPPG(D). For comparison, we built a PPG system based on the same architecture in Fig. \ref{fig:SE system}, and this system is termed PPG(Mono). Two baseline SE systems were implemented: one was based on a  Long-short term memory (LSTM), which was single-directional and fitted the causal implementation, and the other was the Transformer-based SE without incorporating PPG and BPPG information.

\begin{table}[tp!]
\caption{Monophonic and BPC recognition results at different SNR levles}
\label{tab:phone_corr}
\centering
\begin{tabular}{ |c||c|c|c|c| } 
 \hline
 \multicolumn{5}{|c|}{Phoneme/BPCs Accuracy(in \%)} \\
 \hline
 SNR	& Mono  & BPC(D) & BPC(M)  &  BPC(P) \\
 \hline
-5&	10.3&	26.8&	33.3&	27.3\\
0&	16.7&   39.8&	44.1&	37.7\\
5&	24.8&	51.4&	53.9&	46.6\\
10&	32.3&	59.5&	61.2&	55.2\\
15&	42.9&	66.7&	65.9&	60.7\\
 \hline
Avg&	25.4& 48.8&	51.7&	45.5\\
 \hline
\end{tabular}
\end{table}


\begin{table*}[htp]
\caption{Averaged STOI scores for the BPSE with BPPG(M), BPPG(P), and BPPG(D) and monophone PPG. The results obtained using the ground-truth BPPG information were also listed, denoted as GT-PPG(Mono) and BPPG(M). The results of original noisy, and enhanced speech generated by LSTM and Transformer were listed for comparison. The boldface numbers indicate the best results among the testing condition.}
\label{tab:STOI_result}
\centering
\begin{tabular}{ |c||c|c|c|c|c|c|c||c|c| } 
 \hline	
\multirow{2}{*}{SNR} & \multirow{2}{*}{Noisy} & \multirow{2}{*}{LSTM} &  \multirow{2}{*}{Transformer} & \multirow{2}{*}{PPG(Mono)} & \multicolumn{3}{c||}{Broad Phone Class} & \multicolumn{2}{c|}{Ground Truth}\\
 \cline{6-10}
&&&&& 
 BPPG(P) 	&
 BPPG(M)  &
BPPG(D) &
GT-PPG(Mono) &
GT-BPPG(M) \\
 \hline	 
-5&0.595&	0.548&		0.620 &	0.616&	\textbf{0.629}&		\textbf{0.627}&		\textbf{0.628} & 0.679 & \textbf{0.708}\\
0&0.701&	0.686&	0.755&	0.759&	\textbf{0.765}&		\textbf{0.765}&	\textbf{0.763} &0.796 & \textbf{0.808}\\
5&0.800&	0.815&	0.851& 	0.859&	\textbf{0.860}&		\textbf{0.861}&		\textbf{0.859} &0.876 & \textbf{0.879}\\
10&0.880&0.900&	0.912& 0.917&\textbf{0.918}&	\textbf{0.918}&	\textbf{0.917} &0.924 & \textbf{0.925}\\
15&0.935&	0.946&	0.948&	0.950&	\textbf{0.951}& \textbf{0.950}&	\textbf{0.951} &0.953 & \textbf{0.953}\\
\hline
Avg& 	0.782& 0.779&	0.817 	&0.820&	\textbf{0.824}&	\textbf{0.824}&	\textbf{0.823} &0.846 & \textbf{0.855}\\
\hline
\end{tabular}
\end{table*}


\subsection{Experimental result}

\subsubsection{Phonetic/BPC recognition accuracies}
To validate the assumption that additional PPG/BPPG can improve the SE process, we first analyze the correctness of PPG and BPPG generated by the AMs. Given the testing data, the phoneme/BPC accuracies were shown in Table ~\ref{tab:phone_corr}, where monophonic and BPC (based on manner of articulation, place of articulation, and data-driven criteria) recognition results are denoted as "Mono", "BPC(M)", "BPC(P)", and "BPC(D)", respectively.

 From Table~\ref{tab:phone_corr}, we note that the "Mono" system (the overall accuracy is 25.4\%) is very sensitive to noise as compared to the three BPC systems (the overall accuracies range from 45.5\% to 51.7\%). In particular, at the SNR of 0 dB, the accuracy rate of "Mono" drops to 16.7\% while the accuracy rates of BPCs' were around 40\%. These results show that BPC-based systems do not need to distinguish phonemes, some of whose acoustic properties might be highly overlapped, and thus can provide robust recognition performance. To qualitatively analyze the recognition results, Figure~\ref{fig:recognition} shows the spectrograms of sample utterances at two SNR levels (10 dB and 0 dB). The phoneme and BPC(M) recognition results, along with the correct transcription are listed. The results show that at 10 dB-SNR condition, both phoneme and BPC(M) were reliable. However, at the SNR drops to 0 dB, many misrecognized phonemes can be observed, while the BPCs recognition results are still acceptable.
Results of Table~\ref{tab:phone_corr} and Figure~\ref{fig:recognition} suggest that the PPG(Mono) may contain incorrect information that could misguide the SE process.

\subsubsection{Speech enhancement results}
 Table~\ref{tab:STOI_result} presents the STOI scores of the proposed BPSE system and the comparative methods. From the table, we first confirm PPG(Mono) outperforms Transformer baseline in higher SNR conditions, confirming that the monophonic PPG can improve the SE performance. However, we note that PG(Mono) underperforms Transformer baseline in low SNR levels, which may be owing to the incorrect PPG information somehow deteriorates the original SE performance. Next, we note that all of the three BPPG systems provide consistent improvements over Transformer baseline across different SNR levels and outperform the PPG(Mono) systems. The results confirm the robustness of the BPPG that can provide beneficial guide for SE even under very low SNR conditions.
 Among the three BPPG systems, BPPG(M) slightly outperforms the other two systems. The reason is yet to be further investigated.
 
 To further identify the effect of PPG and BPPG, we conducted an additional experiment, where the original phoneme labels from the TIMIT corpus was to get the PPG/BPPG information. More specifically, the input to the AE in the right side of Fig ~\ref{fig:SE system} is an one-hot vector, whose dimension is the number of monophone/BPCs. Here the BPC(M) was used as a representative. This set of results was denoted as Ground Truth (GT-PPG(Mono) and GT-BPPG(M)) in Table~\ref{tab:STOI_result}. From the results of PPG(Mono) and BPPG(M) in the Ground Truth column, BPPG(M) still performed obviously better than PPG(Mono), suggesting that BPPG may be more suitable than PPG to further improve the SE performance.

\begin{figure}[htp]
  \centering
    \subfloat[baseline model and proposed method]{\label{fig:pesq1}\includegraphics[scale=0.18]{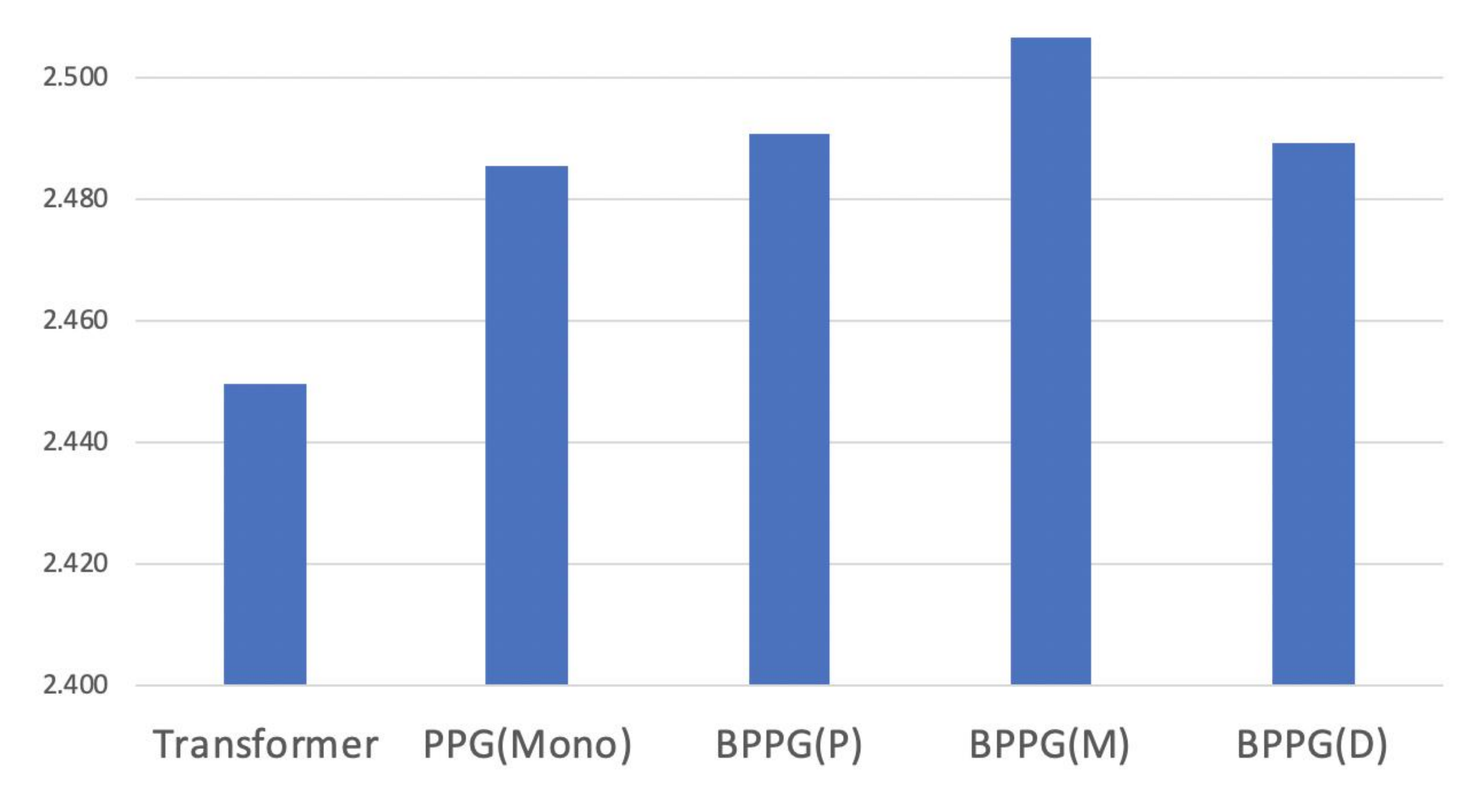}}
    \subfloat[ground truth]{\label{fig:pesq2}\includegraphics[scale=0.133]{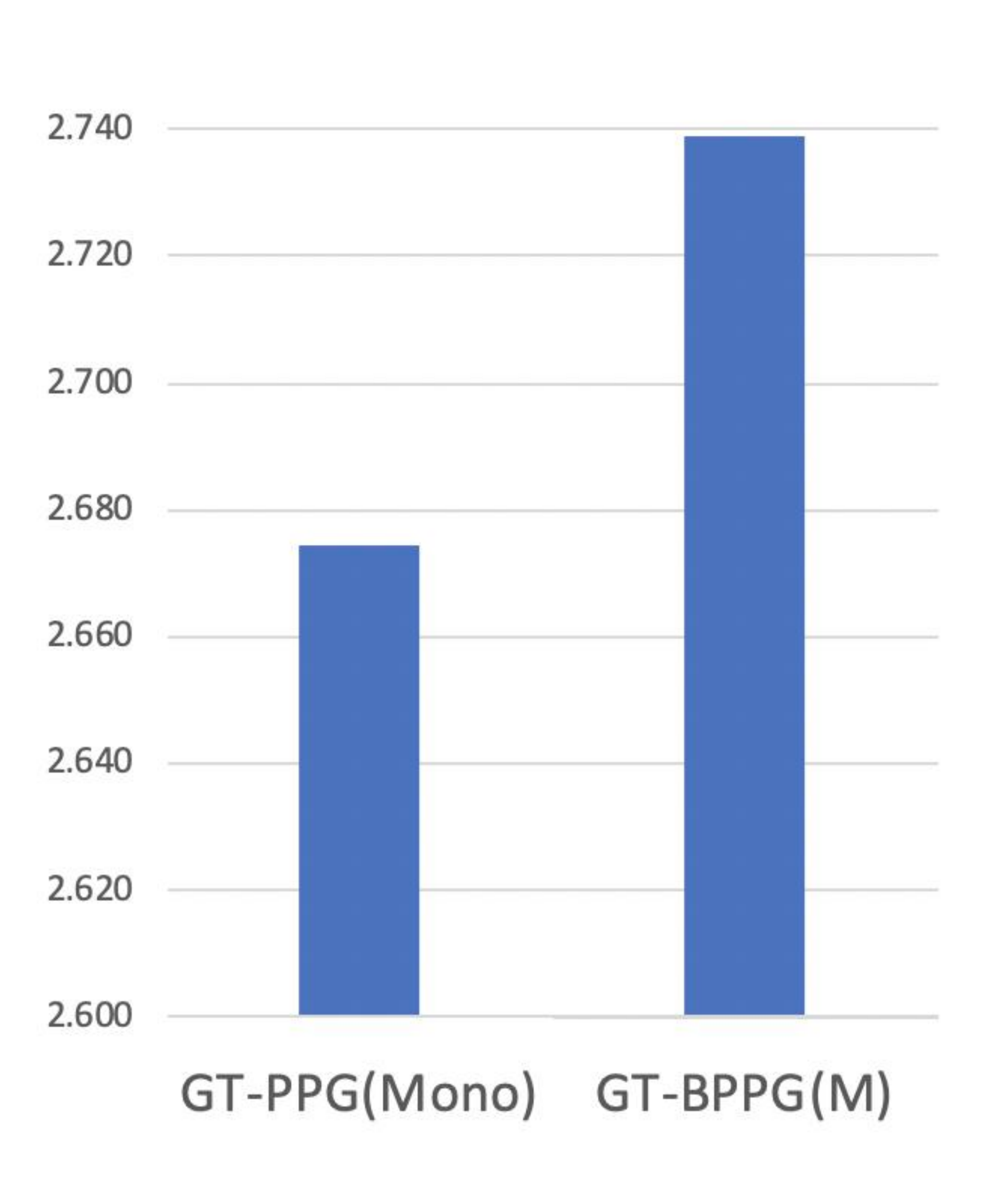}} \\
  \caption{Average PESQ scores baseline model and proposed method}
  \label{fig:PESQ_result}
\end{figure}

 In addition to the STOI scores, Figure~\ref{fig:PESQ_result} presents the average PESQ scores of the SE systems. We note very similar trends to the STOI results as shown in Table~\ref{tab:STOI_result}. All of the three BPPG systems achieve notably better performance in terms of Transformer SE baseline. For the Ground-truth situation, BPPG(M) still yields higher PESQ scores over the PPG(Mono) system.


\section{Conclusion}
In this paper, we proposed to use the BPC information to guide the SE process to achieve better denoising performance. Three clustering criteria were investigated, and the results confirmed that incorporating the BPC information the SE performance can be notably improved over different SNR conditions. The performance even outperforms the SE with monophonic PPG system. The contribution of this work includes: (1) This is the first attempt that incorporates the BPPG to the SE system and obtain promising results. (2) We have verified that both knowledge-based and data-driven criteria can be applied to cluster phonemes into BPCs, both providing beneficial information to the SE system. 
From the ground-truth results, we note that there are still rooms for further improvements. We will further explore how to more effectively use these BPC information for the SE task in the future study.

\bibliographystyle{IEEEtran}

\bibliography{template}


\end{document}